\documentstyle[12pt]{article}
\input tcilatex

\QQQ{Language}{
American English
}

\begin{document}

ALL FOUR GAUGE TRANSFORMATIONS OF THE FERMION

LAGRANGIAN

Gunn A. Quznetsov


Four global gauge transformation of the free fermion Lagrangian is
considered. One of these transformations has S(1) symmetry, other
transformation has SU(2) symmetry, and two others have SU(3) symmetry. It is
supposed, that these transformations determine four grades of the physics
interactions - electromagnetic, weak, gravitational and chromatic.

$\diamondsuit $1.INTRODUCTION

The electromagnetic, weak and strong interactions are conditioned by the
local gauge symmetry of the Lagrangian. The local gauge symmetry can be
gotten from the global gauge symmetry by the substitution of the constant
transformation parameters by the variables. Here I will consider four global
Lagrangian gauge transformations.

Use the natural metric: $\hbar =c=1$.

$\diamondsuit $2.

Let us consider the free fermion Lagrangian:

\begin{equation}
\label{(1)}L=0.5\cdot i\cdot \left( \left( \partial _\mu \overline{\Psi }%
\right) \cdot \gamma ^\mu \cdot \Psi -\overline{\Psi }\cdot \gamma ^\mu
\cdot \left( \partial _\mu \Psi \right) \right) -m\cdot \overline{\Psi }%
\cdot \Psi \text{, } 
\end{equation}

here: $0\leq \mu \leq 3$, $\gamma ^\mu $ are the Clifford matrices:

$$
\gamma ^0=\left[ 
\begin{array}{cccc}
0 & 0 & 1 & 0 \\ 
0 & 0 & 0 & 1 \\ 
1 & 0 & 0 & 0 \\ 
0 & 1 & 0 & 0 
\end{array}
\right] ,\gamma ^1=\left[ 
\begin{array}{cccc}
0 & 0 & 0 & -1 \\ 
0 & 0 & -1 & 0 \\ 
0 & 1 & 0 & 0 \\ 
1 & 0 & 0 & 0 
\end{array}
\right] , 
$$

$$
\gamma ^2=\left[ 
\begin{array}{cccc}
0 & 0 & 0 & i \\ 
0 & 0 & -i & 0 \\ 
0 & -i & 0 & 0 \\ 
i & 0 & 0 & 0 
\end{array}
\right] ,\gamma ^3=\left[ 
\begin{array}{cccc}
0 & 0 & -1 & 0 \\ 
0 & 0 & 0 & 1 \\ 
1 & 0 & 0 & 0 \\ 
0 & -1 & 0 & 0 
\end{array}
\right] , 
$$

$\Psi $ is the spinor, $\overline{\Psi }=\Psi ^{\dagger }\cdot \gamma ^0$ ($%
\overline{\Psi }$ is the adjunct for $\Psi $ spinor).

If $\beta ^\mu =\gamma ^0\cdot \gamma ^\mu $ then

\begin{equation}
\label{(2)}L=0.5\cdot i\cdot \left( \left( \partial _\mu \Psi ^{\dagger
}\right) \cdot \beta ^\mu \cdot \Psi -\Psi ^{\dagger }\cdot \beta ^\mu \cdot
\left( \partial _\mu \Psi \right) \right) -m\cdot \Psi ^{\dagger }\cdot
\gamma ^0\cdot \Psi \text{. } 
\end{equation}

Here:

$$
\beta ^1=\left[ 
\begin{array}{cccc}
0 & 1 & 0 & 0 \\ 
1 & 0 & 0 & 0 \\ 
0 & 0 & 0 & -1 \\ 
0 & 0 & -1 & 0 
\end{array}
\right] ,\beta ^2=\left[ 
\begin{array}{cccc}
0 & -i & 0 & 0 \\ 
i & 0 & 0 & 0 \\ 
0 & 0 & 0 & i \\ 
0 & 0 & -i & 0 
\end{array}
\right] ,\beta ^3=\left[ 
\begin{array}{cccc}
1 & 0 & 0 & 0 \\ 
0 & -1 & 0 & 0 \\ 
0 & 0 & -1 & 0 \\ 
0 & 0 & 0 & 1 
\end{array}
\right] 
$$

Let the spinor $\Psi $ be expressed in the following form:

$$
\Psi =\left| \Psi \right| \cdot \left[ 
\begin{array}{c}
\exp \left( i\cdot g\right) \cdot \cos \left( b\right) \cdot \cos \left(
a\right) \\ 
\exp \left( i\cdot d\right) \cdot \sin \left( b\right) \cdot \cos \left(
a\right) \\ 
\exp \left( i\cdot f\right) \cdot \cos \left( v\right) \cdot \sin \left(
a\right) \\ 
\exp \left( i\cdot q\right) \cdot \sin \left( v\right) \cdot \sin \left(
a\right) 
\end{array}
\right] . 
$$

In this case the probability current vector $\overrightarrow{j}$ has got the
following components:

\begin{equation}
\label{(3)}
\begin{array}{c}
j_1=\Psi ^{\dagger }\cdot \beta ^1\cdot \Psi = \\ 
=\left| \Psi \right| ^2\cdot \left[ \cos ^2\left( a\right) \cdot \sin \left(
2\cdot b\right) \cdot \cos \left( d-g\right) -\sin ^2\left( a\right) \cdot
\sin \left( 2\cdot v\right) \cdot \cos \left( q-f\right) \right] \text{, } 
\end{array}
\end{equation}

\begin{equation}
\label{(4)}
\begin{array}{c}
j_2=\Psi ^{\dagger }\cdot \beta ^2\cdot \Psi = \\ 
=\left| \Psi \right| ^2\cdot \left[ \cos ^2\left( a\right) \cdot \sin \left(
2\cdot b\right) \cdot \sin \left( d-g\right) -\sin ^2\left( a\right) \cdot
\sin \left( 2\cdot v\right) \cdot \sin \left( q-f\right) \right] \text{, } 
\end{array}
\end{equation}

\begin{equation}
\label{(5)}j_3=\Psi ^{\dagger }\cdot \beta ^3\cdot \Psi =\left| \Psi \right|
^2\cdot \left[ \cos ^2\left( a\right) \cdot \cos \left( 2\cdot b\right)
-\sin ^2\left( a\right) \cdot \cos \left( 2\cdot v\right) \right] \text{. } 
\end{equation}

If

\begin{equation}
\label{(6)}\rho =\Psi ^{\dagger }\cdot \Psi \text{, } 
\end{equation}

then $\rho $ is the probability density, i.e. $\stackunder{\left( V\right) }{%
\int \int \int }\rho \left( t\right) \cdot dV$ is the probability to find
the particle with the state function $\Psi $ in the domain $V$ of the
3-dimensional space at the time moment $t$. In this case, $\{\rho ,%
\overrightarrow{j}\}$ is the probability density $3+1$-vector.

If

\begin{equation}
\label{(7)}\overrightarrow{j}=\rho \cdot \overrightarrow{u}\text{, } 
\end{equation}

then $\overrightarrow{u}$ is the average velocity for this particle.

For the left particle (for example, the left neutrino):

$a=\frac \pi 2$,

$$
\Psi _L=\left| \Psi _L\right| \cdot \left[ 
\begin{array}{c}
0 \\ 
0 \\ 
\exp \left( i\cdot f\right) \cdot \cos \left( v\right) \\ 
\exp \left( i\cdot q\right) \cdot \sin \left( v\right) 
\end{array}
\right] 
$$

and from $\left( \ref{(3)}\right) ,\left( \ref{(4)}\right) $, $\left( \ref
{(5)}\right) $, $\left( \ref{(6)}\right) $ and $\left( \ref{(7)}\right) $: $%
u_1^2+u_2^2+u_3^2=1$.Hence, the left particle velocity equals $1$; hence,
the mass of the left particle equals to zero.

The Clifford pentad, which contains the matrices $\gamma ^0$, $\beta ^1$, $%
\beta ^2$, $\beta ^3$, contains the matrix

$$
\beta ^4=\left[ 
\begin{array}{cccc}
0 & 0 & i & 0 \\ 
0 & 0 & 0 & i \\ 
-i & 0 & 0 & 0 \\ 
0 & -i & 0 & 0 
\end{array}
\right] \text{,} 
$$

else. Let us denote:

\begin{equation}
\label{(8)}J_0=\Psi ^{\dagger }\cdot \gamma ^0\cdot \Psi ,J_4=\Psi ^{\dagger
}\cdot \beta ^4\cdot \Psi ,J_0=\rho \cdot V_0,J_4=\rho \cdot V_4\text{.} 
\end{equation}

In this case:

$$
V_0=\sin \left( 2\cdot a\right) \cdot \left[ \cos \left( b\right) \cdot \cos
\left( v\right) \cdot \cos \left( g-f\right) +\sin \left( b\right) \cdot
\sin \left( v\right) \cdot \cos \left( d-q\right) \right] \text{,} 
$$

$$
V_4=\sin \left( 2\cdot a\right) \cdot \left[ \cos \left( b\right) \cdot \cos
\left( v\right) \cdot \sin \left( g-f\right) +\sin \left( b\right) \cdot
\sin \left( v\right) \cdot \sin \left( d-q\right) \right] \text{;} 
$$

and for every particle: $u_1^2+u_2^2+u_3^2+V_0^2+V_4^2=1$. Hence, for the
left particles: $V_0^2+V_4^2=0$.

$\diamondsuit $3.

Lagrangian is invariant for the global gauge transformation:

\begin{equation}
\label{(9)}\Psi \rightarrow \exp \left( i\cdot \alpha \right) \cdot \Psi 
\text{, } 
\end{equation}

here $\alpha $ is the parameter of this gauge transformation. u$_1$, $u_2$, $%
u_3$, $V_0$ and $V_4$ are invariant for this transformation.

Let $U$ be the weak global isospin (SU(2)) transformation with the
eigenvalues $\exp \left( \pm i\cdot \lambda \right) $.

In this case for this transformation eigenvector $\Psi $:

\begin{equation}
\label{(10)}U\Psi =\left| \Psi \right| \cdot \left[ 
\begin{array}{c}
\exp \left( i\cdot g\right) \cdot \cos \left( b\right) \cdot \cos \left(
a\right) \\ 
\exp \left( i\cdot d\right) \cdot \sin \left( b\right) \cdot \cos \left(
a\right) \\ 
\exp \left( i\cdot \lambda \right) \cdot \exp \left( i\cdot f\right) \cdot
\cos \left( v\right) \cdot \sin \left( a\right) \\ 
\exp \left( i\cdot \lambda \right) \cdot \exp \left( i\cdot q\right) \cdot
\sin \left( v\right) \cdot \sin \left( a\right) 
\end{array}
\right] 
\end{equation}

(here ''$\pm $'' is not essential) and for $1\leq \mu \leq 3$:

$$
\left( U\Psi \right) ^{\dagger }\cdot \beta ^\mu \cdot \left( U\Psi \right)
=\Psi ^{\dagger }\cdot \beta ^\mu \cdot \Psi \text{,} 
$$

but for $\mu =0$ and $\mu =4$:

\begin{equation}
\label{(11)}
\begin{array}{c}
\Psi ^{\dagger }\cdot \gamma ^0\cdot \Psi =\left| \Psi \right| ^2\cdot \sin
\left( 2\cdot a\right) \cdot \\ 
\left[ \cos \left( b\right) \cdot \cos \left( v\right) \cdot \cos \left(
g-f-\lambda \right) +\sin \left( b\right) \cdot \sin \left( v\right) \cdot
\cos \left( d-q-\lambda \right) \right] \text{, } 
\end{array}
\end{equation}

\begin{equation}
\label{(12)}
\begin{array}{c}
\Psi ^{\dagger }\cdot \beta ^4\cdot \Psi =\left| \Psi \right| ^2\cdot \sin
\left( 2\cdot a\right) \cdot \\ 
\left[ \cos \left( b\right) \cdot \cos \left( v\right) \cdot \sin \left(
g-f-\lambda \right) +\sin \left( b\right) \cdot \sin \left( v\right) \cdot
\sin \left( d-q-\lambda \right) \right] \text{; } 
\end{array}
\end{equation}

Hence, the Lagrangian $L$ is invariant for this transformation in the all
its member, except the hindmost $\left( \ref{(2)}\right) $:

\begin{equation}
\label{(13)}m\cdot \Psi ^{\dagger }\cdot \gamma ^0\cdot \Psi \text{, } 
\end{equation}

but from $\left( \ref{(11)}\right) $ and $\left( \ref{(12)}\right) $:

\begin{equation}
\label{(14)}m\cdot \left( \left( \Psi ^{\dagger }\cdot \gamma ^0\cdot \Psi
\right) ^2+\left( \Psi ^{\dagger }\cdot \beta ^4\cdot \Psi \right) ^2\right)
^{0.5}\text{ } 
\end{equation}

is invariant for this transformation. If the member $\left( \ref{(13)}%
\right) $ will be substituted by the expression $\left( \ref{(14)}\right) $
in $L$ then $L$ will become invariant for the weak global isospin
transformation.

Let us denote:

$$
O=\left[ 
\begin{array}{cc}
0 & 0 \\ 
0 & 0 
\end{array}
\right] \text{.} 
$$

In this case let the $4\times 4$ matrices of kind:

$$
\left[ 
\begin{array}{cc}
P & O \\ 
O & S 
\end{array}
\right] 
$$

be denoted as the diagonal matrices, and

$$
\left[ 
\begin{array}{cc}
O & P \\ 
S & O 
\end{array}
\right] 
$$

be denoted as the antidiogonal matrices.

Three diagonal members of the Clifford pentad ($\gamma ^0$, $\beta ^1$,$%
\beta ^2$ ,$\beta ^3$ ,$\beta ^4$) define the 3-dimensional space in which $%
u_1$, $u_2$ , $u_3$ are located. The physics objects move in this space. Two
antidiogonal members of this pentad define the 2-dimensional space in which $%
V_0$ and $V_4$ are located. The weak isospin transformation acts in this
space.

$\diamondsuit $4.

Let $\phi _1$ ,$\phi _2$ ,$\phi _3$ be any real numbers (the transformation
parameters); $E$ be the identity $4\times 4$ matrix.

Let

$$
U_1=-i\cdot \beta ^3\cdot \beta ^2,Q_1\left( \phi _1\right) =\cos \left(
\phi _1\right) \cdot E+i\cdot \sin \left( \phi _1\right) \cdot U_1\text{,} 
$$

$$
U_2=-i\cdot \beta ^3\cdot \beta ^1,Q_2\left( \phi _2\right) =\cos \left(
\phi _2\right) \cdot E+i\cdot \sin \left( \phi _2\right) \cdot U_2\text{,} 
$$

$$
U_3=-i\cdot \beta ^1\cdot \beta ^2,Q_3\left( \phi _3\right) =\cos \left(
\phi _3\right) \cdot E+i\cdot \sin \left( \phi _3\right) \cdot U_3\text{.} 
$$

In this case the Lagrangian $L$ is invariant for the following
transformations:

\begin{equation}
\label{(15)}\left\{ 
\begin{array}{c}
\beta ^1\rightarrow \beta ^1 
\text{,} \\ \beta ^2\rightarrow \beta ^2\cdot \cos \left( 2\cdot \phi
_1\right) +\beta ^3\cdot \sin \left( 2\cdot \phi _1\right) 
\text{,} \\ \beta ^3\rightarrow \beta ^3\cdot \cos \left( 2\cdot \phi
_1\right) -\beta ^2\cdot \sin \left( 2\cdot \phi _1\right) 
\text{, } \\ \Psi \rightarrow Q_1\left( \phi _1\right) \cdot \Psi \\ 
\gamma ^0\rightarrow \gamma ^0 
\text{,} \\ \beta ^4\rightarrow \beta ^4\text{.} 
\end{array}
\right\} \text{ } 
\end{equation}

\begin{equation}
\label{(16)}\left\{ 
\begin{array}{c}
\beta ^1\rightarrow \beta ^1\cdot \cos \left( 2\cdot \phi _2\right) +\beta
^3\cdot \sin \left( 2\cdot \phi _2\right) 
\text{,} \\ \beta ^2\rightarrow \beta ^2 
\text{,} \\ \beta ^3\rightarrow \beta ^3\cdot \cos \left( 2\cdot \phi
_2\right) -\beta ^1\cdot \sin \left( 2\cdot \phi _2\right) 
\text{, } \\ \Psi \rightarrow Q_2\left( \phi _2\right) \cdot \Psi \\ 
\gamma ^0\rightarrow \gamma ^0 
\text{,} \\ \beta ^4\rightarrow \beta ^4\text{.} 
\end{array}
\right\} \text{ } 
\end{equation}

\begin{equation}
\label{(17)}\left\{ 
\begin{array}{c}
\beta ^1\rightarrow \beta ^1\cdot \cos \left( 2\cdot \phi _3\right) -\beta
^2\cdot \sin \left( 2\cdot \phi _3\right) 
\text{,} \\ \beta ^2\rightarrow \beta ^2\cdot \cos \left( 2\cdot \phi
_3\right) +\beta ^1\cdot \sin \left( 2\cdot \phi _3\right) 
\text{,} \\ \beta ^3\rightarrow \beta ^3 
\text{, } \\ \Psi \rightarrow Q_3\left( \phi _3\right) \cdot \Psi \\ 
\gamma ^0\rightarrow \gamma ^0 
\text{,} \\ \beta ^4\rightarrow \beta ^4\text{.} 
\end{array}
\right\} \text{ } 
\end{equation}

Hence, these transformations coordinate to the turning in the 3-dimensional
space of $\beta ^1$, $\beta ^2$, $\beta ^3$.

$\diamondsuit $5.

By analogy with $\left( \ref{(15)}\right) $, $\left( \ref{(16)}\right) $, $%
\left( \ref{(17)}\right) $: let

\begin{equation}
\label{(18)}U_0=-i\cdot \gamma ^0\cdot \beta ^4,Q_0\left( \phi \right) =\cos
\left( \phi \right) \cdot E+i\cdot \sin \left( \phi \right) \cdot U_0\text{. 
} 
\end{equation}

In this case, $Q_0$ coordinates to the turning in the 2-dimensional space $%
V_0$, $V_4$ $\left( \ref{(8)}\right) $. Because

$$
Q_0\left( \phi \right) =\left[ 
\begin{array}{cccc}
\exp \left( -i\cdot \phi \right) & 0 & 0 & 0 \\ 
0 & \exp \left( -i\cdot \phi \right) & 0 & 0 \\ 
0 & 0 & \exp \left( i\cdot \phi \right) & 0 \\ 
0 & 0 & 0 & \exp \left( i\cdot \phi \right) 
\end{array}
\right] ` 
$$

then the product of this transformation and the transformation $\left( \ref
{(9)}\right) $ with the parameter $\phi $: $\exp \left( i\cdot \phi \right)
\cdot Q_0\left( \phi \right) $ is the weak global isospin transformation $%
\left( \ref{(10)}\right) $ with the parameter $2\cdot \phi $. Here $%
Q_0\left( \phi \right) $ is the symmetric transformation but $\exp \left(
i\cdot \phi \right) $ disturbs this symmetry.

$\diamondsuit $6.

Let

$$
I=\left[ 
\begin{array}{cc}
1 & 0 \\ 
0 & 1 
\end{array}
\right] 
$$

and

$$
\sigma _1=\left( 
\begin{array}{cc}
0 & 1 \\ 
1 & 0 
\end{array}
\right) ,\sigma _2=\left( 
\begin{array}{cc}
0 & -i \\ 
i & 0 
\end{array}
\right) ,\sigma _3=\left( 
\begin{array}{cc}
1 & 0 \\ 
0 & -1 
\end{array}
\right) 
$$

be the Pauli matrices.

Six Clifford's pentads exists, only:

the red pentad $\zeta $:

$$
\zeta ^1=\left[ 
\begin{array}{cc}
\sigma _1 & O \\ 
O & -\sigma _1 
\end{array}
\right] ,\zeta ^2=\left[ 
\begin{array}{cc}
\sigma _2 & O \\ 
O & \sigma _2 
\end{array}
\right] ,\zeta ^3=\left[ 
\begin{array}{cc}
-\sigma _3 & O \\ 
O & -\sigma _3 
\end{array}
\right] , 
$$

$$
\gamma _\zeta ^0=\left[ 
\begin{array}{cc}
O & -\sigma _1 \\ 
-\sigma _1 & O 
\end{array}
\right] ,\zeta ^4=-i\cdot \left[ 
\begin{array}{cc}
O & \sigma _1 \\ 
-\sigma _1 & O 
\end{array}
\right] ; 
$$

the green pentad $\eta $:

$$
\eta ^1=\left[ 
\begin{array}{cc}
-\sigma _1 & O \\ 
O & -\sigma _1 
\end{array}
\right] ,\eta ^2=\left[ 
\begin{array}{cc}
\sigma _2 & O \\ 
O & -\sigma _2 
\end{array}
\right] ,\eta ^3=\left[ 
\begin{array}{cc}
-\sigma _3 & O \\ 
O & -\sigma _3 
\end{array}
\right] , 
$$

$$
\gamma _\eta ^0=\left[ 
\begin{array}{cc}
O & -\sigma _2 \\ 
-\sigma _2 & O 
\end{array}
\right] ,\eta ^4=i\cdot \left[ 
\begin{array}{cc}
O & \sigma _2 \\ 
-\sigma _2 & O 
\end{array}
\right] ; 
$$

the blue pentad $\theta $:

$$
\theta ^1=\left[ 
\begin{array}{cc}
-\sigma _1 & O \\ 
O & -\sigma _1 
\end{array}
\right] ,\theta ^2=\left[ 
\begin{array}{cc}
\sigma _2 & O \\ 
O & \sigma _2 
\end{array}
\right] ,\theta ^3=\left[ 
\begin{array}{cc}
\sigma _3 & O \\ 
O & -\sigma _3 
\end{array}
\right] , 
$$

$$
\gamma _\theta ^0=\left[ 
\begin{array}{cc}
O & -\sigma _3 \\ 
-\sigma _3 & O 
\end{array}
\right] ,\theta ^4=-i\cdot \left[ 
\begin{array}{cc}
O & \sigma _3 \\ 
-\sigma _3 & O 
\end{array}
\right] ; 
$$

the light pentad $\beta $:

$$
\beta ^1=\left[ 
\begin{array}{cc}
\sigma _1 & O \\ 
O & -\sigma _1 
\end{array}
\right] ,\beta ^2=\left[ 
\begin{array}{cc}
\sigma _2 & O \\ 
O & -\sigma _2 
\end{array}
\right] ,\beta ^3=\left[ 
\begin{array}{cc}
\sigma _3 & O \\ 
O & -\sigma _3 
\end{array}
\right] , 
$$

$$
\gamma ^0=\left[ 
\begin{array}{cc}
O & I \\ 
I & O 
\end{array}
\right] ,\beta ^4=i\cdot \left[ 
\begin{array}{cc}
O & I \\ 
I & O 
\end{array}
\right] ; 
$$

the sweet pentad \underline{$\Delta $}:

$$
\underline{\Delta }^1=\left[ 
\begin{array}{cc}
O & -\sigma _1 \\ 
-\sigma _1 & O 
\end{array}
\right] ,\underline{\Delta }^2=\left[ 
\begin{array}{cc}
O & -\sigma _2 \\ 
-\sigma _2 & O 
\end{array}
\right] ,\underline{\Delta }^3=\left[ 
\begin{array}{cc}
O & -\sigma _3 \\ 
-\sigma _3 & O 
\end{array}
\right] , 
$$

$$
\underline{\Delta }^0=\left[ 
\begin{array}{cc}
-I & O \\ 
O & I 
\end{array}
\right] ,\underline{\Delta }^4=i\cdot \left[ 
\begin{array}{cc}
O & I \\ 
-I & O 
\end{array}
\right] ; 
$$

the bitter pentad \underline{$\Gamma $}:

$$
\underline{\Gamma }^1=i\cdot \left[ 
\begin{array}{cc}
O & -\sigma _1 \\ 
\sigma _1 & O 
\end{array}
\right] ,\underline{\Gamma }^2=i\cdot \left[ 
\begin{array}{cc}
O & -\sigma _2 \\ 
\sigma _2 & O 
\end{array}
\right] ,\underline{\Gamma }^3=i\cdot \left[ 
\begin{array}{cc}
O & -\sigma _3 \\ 
\sigma _3 & O 
\end{array}
\right] , 
$$

$$
\underline{\Gamma }^0=\left[ 
\begin{array}{cc}
-I & O \\ 
O & I 
\end{array}
\right] ,\underline{\Gamma }^4=\left[ 
\begin{array}{cc}
O & I \\ 
I & O 
\end{array}
\right] \text{.} 
$$

The average velocity vector for the sweet pentad has gotten the following
components:

$$
V_0^{\underline{\Delta }}=-\cos \left( 2\cdot a\right) , 
$$

$$
V_1^{\underline{\Delta }}=-\sin \left( 2\cdot a\right) \cdot \left[ \cos
\left( b\right) \cdot \sin \left( v\right) \cos \left( g-q\right) +\sin
\left( b\right) \cdot \cos \left( v\right) \cos \left( d-f\right) \right] , 
$$

$$
V_2^{\underline{\Delta }}=-\sin \left( 2\cdot a\right) \cdot \left[ -\cos
\left( b\right) \cdot \sin \left( v\right) \sin \left( g-q\right) +\sin
\left( b\right) \cdot \cos \left( v\right) \sin \left( d-f\right) \right] , 
$$

$$
V_3^{\underline{\Delta }}=-\sin \left( 2\cdot a\right) \cdot \left[ \cos
\left( b\right) \cdot \cos \left( v\right) \cos \left( g-f\right) -\sin
\left( b\right) \cdot \sin \left( v\right) \cos \left( d-q\right) \right] , 
$$

$$
V_4^{\underline{\Delta }}=-\sin \left( 2\cdot a\right) \cdot \left[ -\cos
\left( b\right) \cdot \cos \left( v\right) \sin \left( g-f\right) -\sin
\left( b\right) \cdot \sin \left( v\right) \sin \left( d-q\right) \right] . 
$$

Therefore, here the antidiogonal matrices $\underline{\Delta }^1$ and $%
\underline{\Delta }^2$ define the 2-dimensional space ($V_1^{\underline{%
\Delta }}$, $V_2^{\underline{\Delta }}$) in which the weak isospin
transformation acts. The antidiogonal matrices $\underline{\Delta }^3$ and $%
\underline{\Delta }^4$ define similar space ($V_3^{\underline{\Delta }}$, $%
V_4^{\underline{\Delta }}$). The sweet pentad is kept a single diagonal
matrix, which defines the one-dimensional space ($V_0^{\underline{\Delta }}$%
) for the moving of the objects.

Like the sweet pentad, the bitter pentad with four antidiogonal matrices and
with single diagonal matrix defines two 2-dimensional spaces, in which the
weak isospin transformation acts, and single one-dimensional space for the
moving of the objects.

$\diamondsuit $7.

Each colored pentad with 3 diagonal matrices and with 2 antidiogonal
matrices, like the light pentad, defines single 2-dimensional space, in
which the weak isospin interaction acts, and defines single 3-dimensional
space for the moving the physics objects.

So in the Lagrangian $\left( \ref{(2)}\right) $ the matrices of the light
pentad are to be replaced by the elements from some colored pentad. Hence
probably, the free fermion Lagrangian has of the form of:

\begin{equation}
\label{(19)}
\begin{array}{c}
L=0.5\cdot i\cdot \left( \left( \partial _\mu \Psi ^{\dagger }\right) \cdot
\kappa ^\mu \cdot \Psi -\Psi ^{\dagger }\cdot \kappa ^\mu \cdot \left(
\partial _\mu \Psi \right) \right) - \\ 
m\cdot \left( \left( \Psi ^{\dagger }\cdot \gamma _\kappa ^0\cdot \Psi
\right) ^2+\left( \Psi ^{\dagger }\cdot \kappa ^4\cdot \Psi \right)
^2\right) ^{0.5} 
\end{array}
\end{equation}

here $\gamma _\kappa ^0$, $\kappa ^1$, $\kappa ^2$, $\kappa ^3$, $\kappa ^4$
are the members or of the light pentad or of a colored pentad.

By analogy with (\ref{(15)}),(\ref{(16)}),(\ref{(17)}): let

\begin{equation}
\label{(20)}\Lambda _\theta =i\cdot \zeta ^0\cdot \eta ^0,G_\theta \left(
\phi _\theta \right) =\cos \left( \phi _\theta \right) \cdot E+i\cdot \sin
\left( \phi _\theta \right) \cdot \Lambda _\theta . 
\end{equation}

In this case the Lagrangian (\ref{(19)}) is invariant for the following
transformation:

$\Psi \rightarrow G_\theta \left( \phi _\theta \right) \cdot \Psi $,

$\zeta ^0\rightarrow \zeta ^0\cdot \cos \left( 2\cdot \phi _\theta \right)
+\eta ^0\cdot \sin \left( 2\cdot \phi _\theta \right) $,

$\eta ^0\rightarrow \eta ^0\cdot \cos \left( 2\cdot \phi _\theta \right)
-\zeta ^0\cdot \sin \left( 2\cdot \phi _\theta \right) $,

$\zeta ^4\rightarrow \zeta ^4\cdot \cos \left( 2\cdot \phi _\theta \right)
-\eta ^4\cdot \sin \left( 2\cdot \phi _\theta \right) $,

$\eta ^4\rightarrow \eta ^4\cdot \cos \left( 2\cdot \phi _\theta \right)
+\zeta ^4\cdot \sin \left( 2\cdot \phi _\theta \right) $,

$\zeta ^1\rightarrow \zeta ^1\cdot \cos \left( 2\cdot \phi _\theta \right)
+\eta ^2\cdot \sin \left( 2\cdot \phi _\theta \right) $,

$\eta ^2\rightarrow \eta ^2\cdot \cos \left( 2\cdot \phi _\theta \right)
-\zeta ^1\cdot \sin \left( 2\cdot \phi _\theta \right) $,

$\zeta ^2\rightarrow \zeta ^2\cdot \cos \left( 2\cdot \phi _\theta \right)
+\eta ^1\cdot \sin \left( 2\cdot \phi _\theta \right) $,

$\eta ^1\rightarrow \eta ^1\cdot \cos \left( 2\cdot \phi _\theta \right)
-\zeta ^2\cdot \sin \left( 2\cdot \phi _\theta \right) $,

$\eta ^3\rightarrow \zeta ^3$,

$\zeta ^3\rightarrow \eta ^3$,

$\theta ^0\rightarrow \theta ^0$,

$\theta ^4\rightarrow \theta ^4$,

$\theta ^3\rightarrow \theta ^3$,

$\theta ^1\rightarrow \theta ^1\cdot \cos \left( 2\cdot \phi _\theta \right)
-\theta ^2\cdot \sin \left( 2\cdot \phi _\theta \right) $,

$\theta ^2\rightarrow \theta ^2\cdot \cos \left( 2\cdot \phi _\theta \right)
+\theta ^1\cdot \sin \left( 2\cdot \phi _\theta \right) $,

$\beta ^0\rightarrow \beta ^0$,

$\beta ^4\rightarrow \beta ^4$,

$\beta ^3\rightarrow \beta ^3$,

$\beta ^1\rightarrow \beta ^1\cdot \cos \left( 2\cdot \phi _\theta \right)
+\beta ^2\cdot \sin \left( 2\cdot \phi _\theta \right) $,

$\beta ^2\rightarrow \beta ^2\cdot \cos \left( 2\cdot \phi _\theta \right)
-\beta ^1\cdot \sin \left( 2\cdot \phi _\theta \right) $.

Therefore, this transformation corresponds to a turning in the space of the
red and the green pentads. Similarly this, the transformations with

\begin{equation}
\label{(21)}\Lambda _\zeta =i\cdot \theta ^0\cdot \eta ^0,G_\zeta \left(
\phi _\zeta \right) =\cos \left( \phi _\zeta \right) \cdot E+i\cdot \sin
\left( \phi _\zeta \right) \cdot \Lambda _\zeta . 
\end{equation}

corresponds to a turning in the space of the green and the blue pentads, and
the transformation with

\begin{equation}
\label{(22)}\Lambda _\eta =i\cdot \theta ^0\cdot \zeta ^0,G_\eta \left( \phi
_\eta \right) =\cos \left( \phi _\eta \right) \cdot E+i\cdot \sin \left(
\phi \eta \right) \cdot \Lambda _\eta . 
\end{equation}

corresponds to a turning in the space of red and blue pentads.

But similar transformation, which corresponds to a turning in the space of a
colored and the light pentads, does not exist.

RESUME

Therefore, the transformation (\ref{(9)}) is the transformation with the
invariance U(1). The transformation (\ref{(10)}) is the transformation with
the invariance SU(2). The transformations (\ref{(15)}), (\ref{(16)}), (\ref
{(17)}) are the transformations with the invariance SU(3) in the space of
the diagonal matrices of the same pentad. The transformations with (\ref
{(20)}), (\ref{(21)}), (\ref{(22)}) are the transformation with invariance
SU(3), too, but in the space of the colored pentads. I'm assume that these
transformations (with (\ref{(20)}), (\ref{(21)}), (\ref{(22)})) determine
the quarks interaction. And because the transformations (\ref{(9)}), (\ref
{(10)}) coordinate to the electroweak interaction then the transformations (%
\ref{(15)}), (\ref{(16)}), (\ref{(17)}) will remain for the gravitation.

That is to say we have got four gauge transformations for four kinds of the
physics interactions: electromagnetic, weak, gravitation, strong.

My other address is: gunn.q@usa.net

\end{document}